# A simple algorithm for decoding both errors and erasures of Reed-Solomon codes

Sergei V. Fedorenko
Department of safety in information systems
St.Petersburg State University of Aerospace Instrumentation
190000, Bolshaya Morskaia, 67, St.Petersburg, Russia

October 2008

**Abstract**

A simple algorithm for decoding both errors and erasures of Reed-Solomon codes is described.

## 1 Introduction

In this paper, the Gao algorithm modification is given. In the author's opinion, the suggested algorithm is the simplest for algebraic codes with short lengths for any implementation.

## 2 Definitions and notations

Let us define the $(n, k, d)$ Reed-Solomon code over $\mathrm{GF}(q)$ with length $n = q-1$, number of information symbols $k$, designed distance $d = n-k+1$, where $q$ is prime power.

The message polynomial of the Reed-Solomon code is

$$M(x) = \sum_{i=0}^{k-1} m_i x^i.$$

The component $c_i$ of the codeword $C(x)$ is computed as

$$c_i = M(\alpha^i), \quad i \in [0, n-1].$$

The received vector is represented as a polynomial

$$R(x) = \sum_{i=0}^{n-1} r_i x^i = C(x) + E(x) = \sum_{i=0}^{n-1} c_i x^i + \sum_{i=0}^{n-1} e_i x^i,$$

where $C(x)$ is the codeword, $E(x)$ is the error vector.

The error vector $E(x)$ has $t$ errors with a set of error positions $\{i_1, i_2, \ldots, i_t\}$. Let us define that $Z_1 = \alpha^{i_1}, Z_2 = \alpha^{i_2}, \ldots, Z_t = \alpha^{i_t}$ are error locations.

The error locator polynomial is

$$W(x) = \prod_{i=1}^{t} (x - Z_i),$$

where $t$ is the number of errors, $Z_i$ is the error location of the error vector $E(x)$.

The error vector $E(x)$ has $l$ erasures with a set of erasure positions $S = \{j_1, j_2, \ldots, j_l\}$. $X_1 = \alpha^{j_1}, X_2 = \alpha^{j_2}, \ldots, X_l = \alpha^{j_l}$ are erasure locations.

The erasure locator polynomial is

$$\Lambda(x) = \prod_{i=1}^{l} (x - X_i),$$

where $l$ is the number of erasures, $X_i$ is the erasure location of the error vector $E(x)$.

The inequality $2t + l < d$ is well known [1].

We construct an interpolating polynomial $T(x)$ such that

$$T(\alpha^i) = r_i, \quad i \in [0, n-1],$$

where $\deg T(x) < n$, and an interpolating polynomial $\mathcal{T}(x)$ such that

$$\mathcal{T}(\alpha^i) = r_i, \quad i \in [0, n-1] \backslash S,$$

where $\deg \mathcal{T}(x) < n - l$.

# 3 Existing algorithms

We describe here two versions of the Gao algorithm [2, 3, 4, 5].

The first version is for decoding errors only. Let $P(x) = W(x)M(x)$. The key equation is

$$\begin{cases} W(x)T(x) \equiv P(x) \mod x^n - 1 \\ \deg W(x) \leq \frac{d-1}{2} \\ \text{maximize } \deg W(x). \end{cases} \quad (1)$$

The asymptotic complexity of this algorithm is $O(n(\log n)^2)$.

The second version is for decoding both errors and erasures. The key equation is

$$\begin{cases} W(x)\mathcal{T}(x) \equiv P(x) \mod \frac{x^n-1}{\Lambda(x)} \\ \deg W(x) \leq \frac{d-l-1}{2} \\ \text{maximize } \deg W(x). \end{cases} \quad (2)$$

The direct computation by this algorithm has complexity $O(n^2)$.

Next, we consider the key equation derivation for the Truong algorithm [6] for decoding both errors and erasures. Let

$$Q(x) = P(x)\Lambda(x) = W(x)M(x)\Lambda(x).$$

From (1) we have

$$W(x)\big((T(x)\Lambda(x))\big) \equiv \big(P(x)\Lambda(x)\big) \mod x^n - 1$$

and the key equation is

$$\begin{cases} W(x)\big((T(x)\Lambda(x))\big) \equiv Q(x) \mod x^n - 1 \\ \deg W(x) \leq \frac{d-l-1}{2} \\ \text{maximize } \deg W(x). \end{cases} \quad (3)$$

The asymptotic complexity of this algorithm coincides with the complexity of decoding algorithms [2, 3, 4, 5].

# 4 Suggested algorithm

We introduce the following lemma.

*Lemma:*
$$T(x) \equiv \mathcal{T}(x) \mod \frac{x^n - 1}{\Lambda(x)}.$$

*Proof:* From Newton's interpolation formula we obtain

$$T(x) = \frac{x^n - 1}{\Lambda(x)} U(x) + \mathcal{T}(x),$$

where $U(x)$ is a polynomial.

From (2) and the lemma we get a new key equation

$$\begin{cases} W(x)T(x) \equiv P(x) \mod \frac{x^n-1}{\Lambda(x)} \\ \deg W(x) \leq \frac{d-l-1}{2} \\ \text{maximize } \deg W(x). \end{cases} \qquad (4)$$

The description of the three algorithms for decoding both errors and erasures is in table 1.

# 5 Conclusion

The suggested algorithm has replaced the computation using Newton's interpolation formula by the fast computation of the discrete Fourier transform. The algorithm complexity is less than the Truong algorithm [6] complexity because the suggested algorithm does not contain some of the intermediate computations.

# References

[1] R. E. Blahut. *Algebraic Codes on Lines, Planes, and Curves: An Engineering Approach.* Cambridge, U.K.: Cambridge University Press, 2008.

[2] A. Shiozaki, "Decoding of redundant residue polynomial codes using Euclid's algorithm," *IEEE Trans. Inf. Theory*, vol. IT–34, no. 5, pp. 1351–1354, Sep. 1988.

Table 1
Algorithms for decoding both errors and erasures

| Step | Gao's algorithm | Truong's algorithm | Suggested algorithm |
|---|---|---|---|
| 0 | — | $\Lambda(x)$ | — |
| 1 | $\mathcal{T}(x)$ | $T(x)$ | $T(x)$ |
| 2a | $\dfrac{x^n - 1}{\Lambda(x)}$ | $T(x)\Lambda(x)$ | $\dfrac{x^n - 1}{\Lambda(x)}$ |
| 2b | $\begin{cases} W(x)\mathcal{T}(x) \equiv P(x) \\ \qquad \mod \frac{x^n-1}{\Lambda(x)} \\ \deg W(x) \leq \frac{d-l-1}{2} \\ \text{maximize } \deg W(x) \end{cases}$ | $\begin{cases} W(x)\big((T(x)\Lambda(x))\big) \equiv Q(x) \\ \qquad \mod x^n - 1 \\ \deg W(x) \leq \frac{d-l-1}{2} \\ \text{maximize } \deg W(x) \end{cases}$ | $\begin{cases} W(x)T(x) \equiv P(x) \\ \qquad \mod \frac{x^n-1}{\Lambda(x)} \\ \deg W(x) \leq \frac{d-l-1}{2} \\ \text{maximize } \deg W(x) \end{cases}$ |
| 3 | $M(x) = \dfrac{P(x)}{W(x)}$ | $M(x) = \dfrac{Q(x)}{W(x)\Lambda(x)}$ | $M(x) = \dfrac{P(x)}{W(x)}$ |
| Complexity | $O(n^2)$ | $O(n(\log n)^2)$ | $O(n(\log n)^2)$ |